\documentstyle[psfig]{EuroPhys}
\input EuroMacr
\input epsf

\def\n{\langle n \rangle}
 \def\gappeq{\mathrel{\rlap {\raise.5ex\hbox{$>$}}
{\lower.5ex\hbox{$\sim$}}}}
\def\lappeq{\mathrel{\rlap{\raise.5ex\hbox{$<$}}
{\lower.5ex\hbox{$\sim$}}}}

\begin{document}
\shorttitle{Non-equilibrium wetting}
\title{ \large \bf Stochastic theory of non-equilibrium wetting}
\author{F. de los Santos $^{1,3}$, M.M. Telo da Gama $^1$,
 and Miguel A. Mu\~noz $^{2,3}$}
\institute{
$^{1}$ Departamento de F{\'\i}sica da Faculdade de Ci{\^e}ncias e Centro
 de F{\'\i}sica da Mat{\'e}ria Condensada da Universidade de
Lisboa, Avenida Professor Gama Pinto, 2, P-1643-003 Lisboa Codex,
Portugal \\
$^{2}$ Depto. de E. y F\'\i sica de la Materia,
Universidad de Granada, 18071 Granada, Spain     \\
$^{3}$ Institute Carlos I for Theoretical and Computational Physics,
University of Granada, 18071 Granada, Spain.
 }
\rec{}{}
\pacs{
\Pacs{05}{70.Fh}{Phase transitions general studies}
\Pacs{68}{45.Gd}{Depinning}
}
\maketitle

\begin{abstract}
We study a Langevin equation describing non-equilibrium
depinning and wetting transitions. Attention is focused on short-ranged
attractive substrate-interface potentials.
We confirm the existence of first order depinning transitions,
in the temperature-chemical potential diagram, and a tricritical
point beyond which the transition becomes a non-equilibrium
complete wetting transition.
The coexistence of pinned and depinned interfaces occurs over a finite
area,
in line with other non-equilibrium systems that exhibit first order
transitions.
In addition, we find two types of phase coexistence, one
of which is characterized by spatio-temporal intermittency (STI). A finite
size analysis of the depinning time is used to characterize the different
coexisting regimes.
Finally, a stationary distribution of characteristic triangles or
facets was shown to be responsible for the structure of the STI phase.
\end{abstract}
\vspace{4pt}

Consider a bulk phase ($\alpha$) in contact with a substrate.
Wetting occurs when a macroscopic layer of a different, coexisting,
bulk phase ($\beta$) is adsorbed at the substrate.
The wetting transition is characterized by the divergence of the
wetting layer thickness, {\em i.e.\/} the distance between the
substrate and the $\alpha \beta$ interface.
Equilibrium wetting is a problem of outmost importance. It
has been observed experimentally and thoroughly investigated
using (among other techniques) interface displacement
models \cite{rev-wet}.
In these models one considers a function $h({\bf x})$ representing
the local departure of the interface from a reference plane ($h=0$) and
constructs an effective interface Hamiltonian, ${\cal H}(h)$.
Using reasonable approximations one arrives at the standard form
\cite{rev-wet},
\begin{equation}
 {\cal H}(h)=\int_0^\infty d{\bf x}\Big[ {1 \over 2} D (\nabla h)^2 +V(h)
\Big],
\label{ham}
\end{equation}
where $D$ is the interfacial tension of the $\alpha \beta$
interface (or the interfacial stiffness for anisotropic media) $V(h)$ is
the external potential
accounting for the net interaction between the substrate
and the $\alpha \beta$ interface and ${\bf x}$ is a $d$-dimensional vector.
For large values of $h$ at bulk coexistence,
the interface potential of a system with short-range forces
vanishes exponentially and one writes \cite{rev-wet}
\begin{equation}
 V(h)=b(T)e^{-h}+ce^{-2h}
\end{equation}
where $h$ is measured in units of the bulk correlation length $\xi$,
$b\equiv b(T)$ vanishes linearly with $T-T_W$, where $T_W$ is the
(mean-field) wetting transition temperature and $c>0$.
At sufficiently low temperatures, $b<0$, the potential $V$ binds the
$\alpha \beta$ interface, that is, the equilibrium thickness of the
wetting layer $\langle h \rangle$ is finite.
As the temperature is raised, the potential becomes less attractive and
at bulk coexistence it no longer binds the interface;
$\langle h \rangle$ diverges.
In this regime, a linear term $\mu h$, where
$\mu$ is the chemical potential difference
between the $\alpha$ and $\beta$ phases, may be added to $V(h)$ to
study complete wetting \cite{rev-wet}.

A dynamic generalization of the equilibrium wetting models
is given by the following Langevin equation
\begin{equation}
\partial_t h({\bf x},t) =
D \nabla^2h -\partial V / \partial h + \eta({\bf x},t) ,
\label{eweq}
\end{equation}
where $\eta$ is a zero-mean Gaussian white noise.
This was introduced by Lipowsky \cite{long}, and describes the
relaxation of $h$ towards its equilibrium distribution.
In this context, $\mu$ plays the role of an external force
acting on the interface. In the absence of the substrate, $\mu=0$ guarantees
that the mean velocity of the interface is zero
regardless of its average position, $\langle h \rangle$, as
required by bulk coexistence.

The dynamic model is readily generalized to
non-equilibrium interfacial processes, {\em e.g.\/}, crystal growth,
atomic beam epitaxy, etc., where thermal equilibrium does not apply. An
effective non-equilibrium interfacial model consists of a
Kardar-Parisi-Zhang (KPZ) equation \cite{reviews} in the presence of a
substrate. Consider the free interface that is solution
of the KPZ equation. A depinning transition occurs at a chemical
potential, $\mu=\mu_c$, that depends on the strength of the KPZ
non-linearity. The non-equilibrium analogous of complete wetting
corresponds to a depinning transition, that occurs at $\mu=\mu_c$, in the
presence of the substrate.

A discrete model that exhibits this type of transitions
was introduced in \cite{haye1}. Interfacial growth was analyzed for
a model with adsorption/desorption rules that violate, in general,
detailed balance. It was shown \cite{haye1} that the corresponding
effective interface model is a KPZ equation in the presence of a repulsive
external
potential. Under equilibrium conditions, the KPZ nonlinearity vanishes
\cite{haye1} and the model reduces to that of equilibrium wetting
\cite{long}.
The difference between the adsorption and desorption rates plays the role
of a driving force (analogous to the chemical potential difference,
$\mu$),
and controls the mean velocity of the interface. For repulsive substrates,
a depinning or complete wetting transition was shown to occur at $\mu =
\mu_c$, the value where the mean velocity of the free interface vanishes.
Under equilibrium conditions the KPZ non-linearity vanishes and a
complete wetting transition occurs at $\mu_c=0$.
An earlier study of non-equilibrium complete wetting was based on the KPZ
equation \cite{reviews} in the presence of a {\it hard wall} \cite{mn}.
It was shown \cite{mn,haye1}, that this non-equilibrium transition
exhibits scaling properties characterized by the KPZ exponents, in
addition to those associated with the pinned phase, that are described by
new critical exponents.
It was also established that there are two universality classes
determined by the sign of the KPZ non-linearity \cite{mn,haye1}.

Subsequently, the effect of short-range attractive substrates was studied
for the discrete model \cite{haye2}. It was found, that for sufficiently
attractive substrate potentials, the depinning transition becomes
first-order and that, by contrast with equilibrium systems, at a given
temperature there is a finite range of chemical potentials for which phase
coexistence occurs. These results were confirmed by Giada and Marsili
\cite{marsili} using a KPZ equation in the presence of an attractive
substrate. In contrast with the transitions that occur for repulsive
substrates the first-order transition is not driven by the
(attractive) substrate.
Here we carry out extensive studies of a similar KPZ equation, using
mean field techniques and numerical simulations for repulsive and
attractive substrate potentials. Some of our results for the first-order
transition are similar to those of \cite{marsili,raul} while others
are new, namely the existence of spatio temporal intermittency (STI),
facets, and two distinct regimes in the coexistence region. We argue that
the continuous transitions that occur for repulsive substrates are
non-equilibrium complete wetting; in addition, we discuss the
nature of the first order depinning transitions that occur in the presence
of attractive substrates.

{\it The model --}
We consider a KPZ equation with a substrate potential
given by $V(h)= -(a+1) h + b  e^{-h}+c e^{-2h}/2$
with $c>0$ \cite{marsili},
\begin{equation}
\partial_t h({\bf x},t)= D\Big[\nabla^2 h-(\nabla h)^2\Big]
-{\partial V(h)\over \partial h} + \eta.
\label{kpzyv}
\end{equation}
$a+1$ acts as a chemical potential difference
between the two phases and $b$ is the temperature measured with respect to
the (mean-field) equilibrium wetting transition temperature.
Thus for $b>0$ this equation describes non-equilibrium complete
wetting. For positive values of the KPZ non-linearity,
it is easy to show that only second order transitions
(in the multiplicative noise, MN, universality class) occur.
 Performing a Cole-Hopf transformation,
$h({\bf x},t)=-\ln n({\bf x},t)$, and using Ito calculus
the last equation transforms into
\begin{equation}
\partial_t n=D\nabla^2 n-{\partial V(n)/ \partial n} + n \eta,
\label{mn}
\end{equation}
with $V(n)=an^2/2+ bn^3/3 +c n^4/4$, that describes the
interfacial problem as a diffusion-like equation with
multiplicative noise. In general, any potential of the form
$an^2/2+b n^{p+2}/(p+2)+c n^{2p+2}/(2p+2)$ with $p>0$ yields an
equivalent effective Hamiltonian, since when the last equation
is written in terms of $h$, the parameter $p$ may be set to one through a
redefinition of the height variable.
The case $p=2$ (with fixed $b>0$) has been studied in \cite{raul}
in the context of stochastic STI.
Note that in terms of $n$ the two phases are
absorbing (depinned) and active (pinned)
respectively, and the critical point signals a transition
to an absorbing state \cite{tipos}.

{\it Mean field --}
In this section we review the mean field analysis of the MN equation.
Most of this was presented in \cite{muller,marsili,walter}; here, we
outline the main results. First the MN equation is discretized.
In the discretized Laplacian,
the sum of the nearest-neighbor field values
is replaced by the average value of the field, $\langle n \rangle$,
and a closed Fokker-Planck equation is obtained for
$P(n,t,\n)$.
This equation is solved self-consistently, by equating the
average value of $n$, $F(\n)$,
with $\langle n \rangle$. Except for the limiting cases
$D=0$ and $D=\infty$ where analytical solutions exist,
\cite{walter,raul,marsili}
the self-consistency equation is solved numerically.

Setting (without loss of generality)
$D=2,c=1.5$ and $p=2$,
three different regimes were found for $b<0$ (Fig.1):
(i) Depinned phase: one stable solution, $n=0$
(dash-dotted line);
(ii) Coexistence: two stable solutions and
one unstable (dashed line);
(iii) Pinned phase: one unstable solution, $\n
=0$, and one stable, $\n \not= 0$ (solid line). For positive values
of $b$ only regimes (i) and (iii) are found, consistent with the
continuous nature of the complete wetting transition.
The associated phase diagram is depicted in Fig. 1.
The solid line is the continuous phase boundary between depinned and
pinned phases. In the region delimited by the dashed lines,
both of which correspond to first-order boundaries, the
depinned and pinned phases coexist as stationary solutions of the
dynamical equation. The three lines meet at the (mean-field) tricritical
point $a=0$ and $b=0$.

\begin{figure}[t]
\epsfxsize=12cm
\epsffile{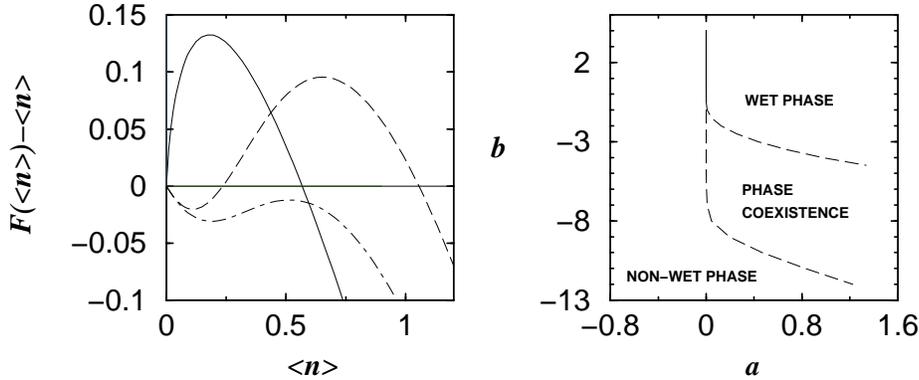}
\caption{Typical solutions of the equation $F(\n)=
\n$ for different values of $a$,
 and mean-field phase diagram.}
\end{figure}

{\it Beyond mean-field theory --}
We have verified by means of extensive numerical simulations
that a phase diagram analogous to that of Fig. 1 survives the
effects of fluctuations.
The phase boundaries were located by fixing $b$ and tuning $a$.
For large (positive) values of $b$ the transition is found to be
continuous. Mean-field results are expected to hold at this transition
above the upper critical dimension $d_c=2$ \cite{mn};
in fact, if $b>0$ and $d>2$, the term $n^{2p+2}$
is irrelevant in the renormalization group sense,
and we are left with the {\it multiplicative noise} equation studied
in \cite{mn,tipos}.
For completeness, we have calculated the order parameter
critical exponent, $\beta$, in dimensions above and below $d_c$.
In $d=1$ ($d=3$) we found $\beta =1.65 \pm 0.05$ ($\beta=0.96 \pm 0.05$)
in agreement, within error bars, with the prediction for the MN class
\cite{mn,walter}.

Let us now focus on the first order transition.
We fixed (without loss of generality) $p=2$, $b=-4$, $c=1.5$, $D=2$.
For these values of the parameters a first order transition is expected
at the mean field level. In order to establish the phase boundaries,
we performed numerical simulations
for one-dimensional systems with system sizes up to $L=2000$,
starting the runs with two different types of initial conditions (IC):
(1) Pinned interfaces with typical values of $h$ close to the wall,
or equivalently $n=1$, {\em i.e.\/}, in the active regime.
We refer to this as `pinned' IC.
(2) Depinned interfaces with large values of $h$
(small values of $n$, close to the absorbing state $n=0$).
In general, an interface is considered detached from the substrate
 when all the values of $h({\bf x})$ are positive
($n({\bf x}) < 1, ~~\forall {\bf x}$).
First, we have verified that the depinned phase is stable for values
of $a$ larger than $a \sim 0$.
The behavior for pinned IC, however, is
much harder to analyze due to finite size effects: for any finite size
the only stationary state is the depinned (absorbing) one.
Thus, in order to establish the stability of the pinned phase,
one has to perform a finite size analysis
of the characteristic time needed
for a system with pinned IC to depin \cite{haye2}.
We have measured these times in two different
ways: (i) the time $\tau$ that characterizes the
exponential decay of $\langle n(t) \rangle$, averaged
over all runs ($ \forall a$), after a short transient;
and (ii) the time needed for the last site in the interface to
detach ($n({\bf x}) < 1, ~~\forall {\bf x}$).
Both lead to the same qualitative results.
\begin{figure}[ht]
\epsfxsize=12cm
\epsffile{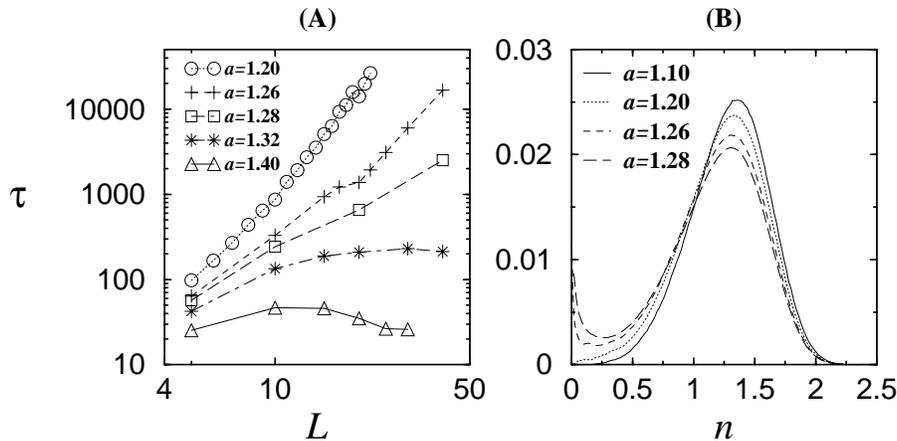}
\caption{Characteristic times for the decay of $\langle n(t) \rangle$
and single
site (unnormalized) pdf for different values of $a$.}
\label{fig2}
\end{figure}
 We observed three regimes:

(1) For $a \lappeq 1.22$
there is an exponential growth of
$\tau$ as $L \rightarrow \infty$, implying that the
pinned state is stable in the thermodynamic limit.
As the depinned phase is stable for $a \gappeq 0 $
the system exhibits coexistence of the depinned and pinned phases
in the range $0 \lappeq a \lappeq 1.22 $ in analogy with other 
nonequilibrium systems \cite{Toom}.
Note that owing to the huge characteristic times in this regime,
obtaining good statistics, even for relatively small systems, is
prohibitive.
Within error-bars, however, the overall behavior is compatible with an
exponential.

(2) In the range $ \approx 1.22 \lappeq a \lappeq 1.3$
an approximate power-law dependence is found, at least for the
accessible system sizes (see Fig. 2A). The pinned phase is stable
 as $\tau$ tends to infinity, in the infinite size limit. Note also the
presence of a step separating two different regimes (see curve for $a=1.26$
 in Fig. 2); this step has the same origin as the maximum in the curves
of the third regime to be discussed below.
Much larger simulations, unaccessible to the available computing
power, are required in order to establish whether $(i)$ 
these power laws describe
the true asymptotic behavior, $(ii)$ for very large systems they become
exponentials and $(iii)$ the power-law region shrinks to a line. 
The CPU time required to measure $\tau$ accurately,
for large $L$, is huge as the distribution of detaching times is
very broad. 

(3) For $a \gappeq 1.3$, $\tau$ tends to a constant, {\em i.e.\/}
in the thermodynamic limit, the system is depinned in a finite time.
Note in Fig. 2 the rather unusual dependence on the system size.
 For instance, for $a=1.4$ the curve has a maximum around
$L=11$, and then decays monotonically. This means that the interface is
more easily detached in a large system than in a smaller one,
which is counterintuitive. This can be rationalized by considering on the
 one hand a depinning `zipper' mechanism: once a site is detached it pulls
out its neighbors, and they in turn pull out their neighbors,
and so on, until (if the system is small enough)
the whole interface is depinned (in a time which grows linearly with $L$).
The existence of this mechanism was verified numerically.
On the other hand, the larger the system size $L$, the larger  the
probability of a fluctuation that detaches a single site. The competition
of these two mechanisms is responsible for the non-monotonic behavior of
the characteristic times in the third regime (as well as for the step in
the previous regime).

To clarify the nature of the second regime
we performed further analysis. Fig. 3A, depicts a typical
spatio-temporal snapshot of the field $n$, for $a=1.28$.
Patterns characteristic of STI \cite{raul} are observed in
this regime (and only in this regime). The pattern is characterized by the
simultaneous presence of pinned and depinned patches, and the distribution
of the latter is rather broad. This regime was unnoticed in previous
studies of non-equilibrium depinning transitions \cite{muller,marsili}.
\begin{figure}[ht]
\epsfxsize=13cm
\epsffile{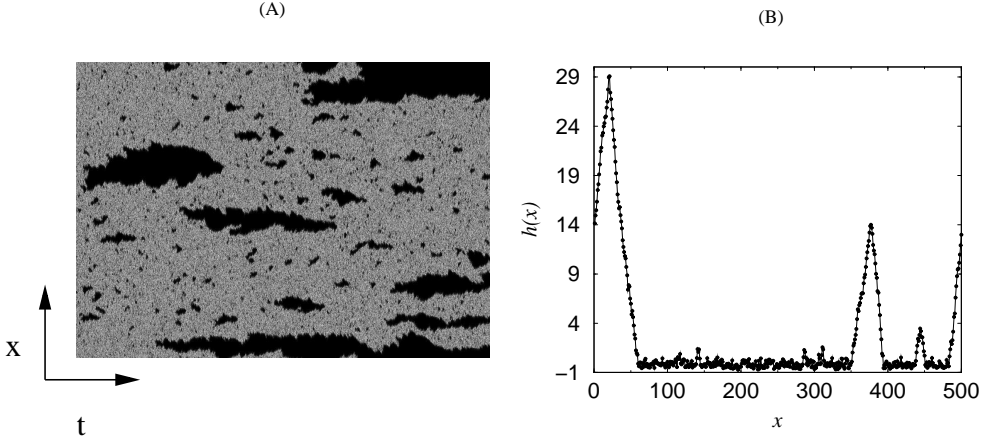}
\caption{A. Snapshot for $a=1.28$;  (dark = depinned, $n <1~$;
grey = pinned, $n>1$). Islands of very different sizes
of the minority phase are ubiquitous.
The $1000$ depicted time slices were taken at intervals of $50$
time units. $L=500$.
B. Typical configuration of $h$ corresponding to the time slice 400, 
marked with a line in the previous snapshot.}
\label{fig3}
\end{figure}
STI occurs for a narrow set of parameters only and not generically in
the pinned phase, contrary to a previous suggestion \cite{raul}. In fact,
the different exponential and power-law regimes, in the
coexistence region, was unnoticed in that work \cite{raul}. Roughly
speaking the STI
regime, where statistically stationary patches of depinned
and pinned regions are seen in generic configurations,
coincides with the second or power-law regime reported in this work.

In order to quantify the distinction between the STI
and the non-STI (globally) pinned phases, we have studied the single
site  probability distribution function (ss-pdf).
The ss-pdf was introduced in \cite{tipos} as a suitable tool
to distinguish between different absorbing states.
Instead of averaging $n(t)$ over all runs, as before, we
average only over the pinned (surviving) runs starting from pinned IC.
This yields non-trivial stationary values of $\n$ in the first
two regimes, while in the third the only stationary state is $n=0$.
For small values of $a$ ($a \lappeq 1.22$) the stationary histogram is
bell-shaped (see Fig. 2B). For $a$ slightly larger than $1.22$
({\em e.g.\/} $1.26$ or $1.28$)
the ss-pdf retains its bell-shape while developing
a second peak around $n=0$. The relative height of the two peaks changes
continuously as $a$ increases.
At a certain value of $a$, signaling the first order
transition, the peak at zero $n$ increases dramatically and the pinned
phase becomes unstable. Physically, a large number of depinned islands is
generated that eventually coalesce and depin the interface.
Consequently, the STI is restricted to a narrow region below the first
order boundary where the two peaks of the stationary ss-pdf co-exist, or
in other words, where detached patches are generated but are suppressed in
times comparable to the typical coalescence time \cite{haye2}. A
typical configuration of $h$ in the
STI regime ($a=1.28$) is shown in Fig. 3B.
Note the presence of triangles analogous
to those described in \cite{haye2};
in fact, the mechanism responsible for these
triangular facets, is similar to that described
in \cite{haye2}, though here, there is no height-gradient
constraint. The (average) slope of each triangle, $s$,
is a constant (modulo fluctuations), and may be determined from
the condition $\lambda_R s^2 = a+1$, where the renormalized value of
the non-linear coefficient $\lambda_R$ is measurable by globally tilting the
interface \cite{reviews,Jeong}.
Similar triangles, have been reported for
the Frenkel-Konrontova equation \cite{FK} and for the KPZ
dynamics with negative non-linearities in the presence of quenched
disorder \cite{Jeong}.


Summing  up, we have studied a general Langevin equation describing
non-equilibrium interfacial phenomena, for short-range interactions
between the interface and the substrate.
We have found, that the non-equilibrium transitions may be continuous or
first order. The former are complete wetting transitions driven by the
repulsive substrate while the latter correspond to the non-equilibrium
growth of an interface driven by the chemical potential difference.
In the first-order regime there is coexistence
between pinned and depinned interfaces for a finite
range of parameters.
Thus, the approach to the critical wetting transition
($a=b=0$ within mean-field) may be realized along
different paths, within the finite coexistence region
(delimited by dashed lines in the mean-field diagram
of Fig. 1).
Moreover, within  the coexistence region there
are two sub-regimes: one exhibiting STI (characterized by a two-peaked
 ss-pdf) and one with a more standard pinned phase
(with no STI and a bell-shaped ss-pdf). In the former the
characteristic times grow with $L$ (roughly) in an algebraic fashion,
while in the latter the growth is exponential. Finally, we have reported
on the existence of triangular structures or facets in one-dimensional
interfaces.

Several important issues remain open and should be addressed before a
connection with experiments is made.
The key task is that of developing criteria
to determine whether the KPZ non-linear term should be included in the
effective interface Hamiltonian of a given system. The KPZ
nonlinearity describes lateral growth, a mechanism
that is unlikely to be relevant for fluid interfaces, but may determine
the (anisotropic) growth behavior observed in crystals \cite{Villain} at
least in systems with short-ranged interactions.
In addition, it remains to be checked whether the phenomenology described
here survives in higher dimensions and/or the presence of long-ranged
interactions. Finally, the crossover
between non-equilibrium wetting transitions and equilibrium
wetting, i.e., the weak coupling limit of the
KPZ equation in the presence of a substrate, remains a challenging
theoretical problem. By analogy with short-ranged equilibrium wetting,
it is expected that the study of non-equilibrium critical wetting
will reveal the existence of new universality classes, that may
depend on the path within the finite coexistence region.

{\it Acknowledgments --}
We acknowledge partial support
from the European Network contracts
ERBFMRXCT980183 and  ERBFMRXCT980171,
the Spanish DGESIC project PB97-0842 and
a running grant of the FCT (Portugal) under their Pluriannual Programme.


\end{document}